\documentclass[amsmath,amssymb,superscriptaddress]{revtex4}
\usepackage{graphics,epsfig}
\usepackage{epstopdf}
\usepackage{graphicx}
\usepackage{dcolumn}
\usepackage{amsmath}
\newcommand*{\car}{{\cal R}}
\usepackage[colorlinks=true, citecolor=blue, urlcolor = blue, linkcolor= red, bookmarks=true]{hyperref}
\begin{document}
	\title{Clouds of strings in $4D$ Einstein-Gauss-Bonnet black holes}
	\author{Dharm Veer Singh}
	\email{veerdsingh@gmail.com}
	\affiliation{Department of Physics, Institute of Applied Science and Humanities, 
		G.L.A University, Mathura, 281406 India.}
	\author{Sushant G. Ghosh}
	\email{sghosh2@jmi.ac.in, sgghosh@gmaill.com}
	\affiliation{Centre for Theoretical Physics, Jamia Millia Islamia, New Delhi 110025,India}
	\affiliation{Astrophysics and Cosmology Research Unit, School of Mathematics, Statistics and Computer Science, University of KwaZulu-Natal, Private Bag X54001, Durban 4000, South Africa}
	\author{Sunil D. Maharaj }\email{maharaj@ukzn.ac.za}
	\affiliation{Astrophysics and Cosmology Research Unit,	School of Mathematics, Statistics and Computer Science,
		University of KwaZulu-Natal, Private Bag X54001, Durban 4000, South Africa}

	\begin{abstract}
Recently there has been significant interest in regularizing, a $ D \to 4 $ limit, of EGB gravity, and the resulting regularized $4D$ EGB gravities have nontrivial gravitational dynamics  - namely the $4D$ EGB gravity. We present an exact charged black hole solution to the $4D$ EGB gravity surrounded by clouds of string (CS) and also analyze their thermodynamic properties. Owing to the corrected black hole due to the background CS, the thermodynamic quantities have also been corrected except for the entropy, which remains unaffected by a CS background. However, as a result of the   $4D$ EGB theory, the Bekenstein-Hawking area law turns out to be corrected by a logarithmic area term. The heat capacity $C_+$ diverges at a critical radius $r=r_C$,  where incidentally the temperature has a maximum, and $C_+ > 0$ for $r_+ < r_C$ allowing the smaller black hole to become locally stable. Interestingly,  due to the surrounding cloud of strings, we have phase transition from globally thermodynamically small stable black holes with negative free energy ($F_{+}<0$) to large unstable black holes.   Our solution can also be identified as a $4D$ monopole-charged EGB black hole.  Our results demonstrate that the Hawking’s evaporation leads to a thermodynamically stable remnant with vanishing temperature. We regain results of spherically symmetric black hole solutions of general relativity and that of $4D$ EGB, respectively, in the limits $\alpha \to 0$ and $a=0$.
	\end{abstract} 
	\maketitle

\section{\label{sec:level1}Introduction}
Lovelock gravity is one of the natural generalizations of Einstein's general relativity (GR) to higher dimensions (HD), introduced by David Lovelock \cite{dll}, the action of which contains series of higher order curvature terms.  While well motivated, higher curvature gravities introduce a number of hurdles making their investigation difficult, e.g., the equations of motion, in such theory, are  fourth order or higher, and  linear perturbations  disclose that the graviton is a ghost.
However, Lovelock theories are distinct, among general higher curvature theories, in having field equations involving not more than second derivatives of the metric and theories are free from many of the problems that affect other  higher derivative gravity theories.   In the Lovelock action \cite{dll}, apart from the cosmological constant  ($\Lambda$) and Einstein GR scalar ($\mathcal{R}$)  as the first two terms, the third  term is a  combination of the second order curvature term, namely Gauss-Bonnet  \cite{Lanczos:1938sf}.  The simplest case of Lovelock theory that departs from GR is the
EGB theory in which  the Einstein-Hilbert action is supplemented with the quadratic curvature GB  term given by 
	\begin{equation}
	{\cal L}_{GB}=\car^2-4\car_{cd}\car^{cd}+\car_{cdef}\car^{cdef}, 
	\label{GB-Lagrangian}
	\end{equation}
where a new  constant $\alpha$ that can be identified as the inverse of the string tension. 
This special case of  Lovelock gravity has received the most significant attention and is called Einstein-Gauss-Bonnet  (EGB) gravity \cite{Lanczos:1938sf}, which naturally appears in the low energy effective action of heterotic string theory \cite{Gross}.    The spherically symmetric static black hole  solution for the EGB gravity was  first  obtained by Boulware and Deser \cite{bd}, A cascade of subsequent interesting work analysed black hole solutions in EGB gravity \cite{egb2,ipn,noc,Anninos:2008sj,Nojiri:2001ae} for various sources including
clouds of strings (CS) \cite{hr,Ghosh:2014pga,Ghosh:2014dqa,Lee:2014dha,Graca:2016cbd}.  Some of the EGB black holes have been shown to exhibit Hawking-Page type transitions in AdS spacetimes \cite{ipn,noc,Anninos:2008sj,Nojiri:2001ae}, CS models \cite{Ghosh:2014pga, hr}, other backgrounds \cite{sgg1} and also for regular black holes \cite{Ghosh1:2018bxg}. 
	
As a HD  member of Einstein's GR family, EGB gravity allows us to explore several conceptual
issues in a broader setup. However, the GB invariant  is a topological invariant  in $4D$ as its contribution to all components of Einstein's equations are in fact proportional to $(D-4)$, and one requires $D\geq 5$ for non-trivial gravitational dynamics.   However, it was shown that by rescaling the GB coupling constant as $\alpha\to \alpha/(D-4)$ the GB invariant, in the limit $D\to4$ when  finding equations, makes a non-trivial contribution to the gravitational dynamics even in  $D=4$ \cite{gla}.  The theory preserves  the number of degrees of freedom and remains free from Ostrogradsky instability \cite{gla}.  Further,  this extension of  Einstein's gravity bypasses conditions of Lovelock's theorem \cite{Lovelock:1972vz} and   For definiteness, the effective gravity is called the   $4D$ EGB theory, which admit spherically symmetric black hole solutions generalizing the Schwarzschild black holes and is also free from  singularity \cite{gla}.    It is argued, without an explicit proof, that a physical observer could never reach this curvature singularity given
the repulsive effect of gravity at short distances \cite{gla}.  However,  considering the geodesic equations, this claim was refuted by  Arrechea {\it et al.} \cite{Arrechea:2020evj}.  They explicitly showed that the infalling
particle starts at rest  will reach the singularity with zero velocity
as  attractive and repulsive effects compensate each other along the trajectory of the
particle  \cite{Arrechea:2020evj}. 

 Further, we would like to draw the attention of the reader that black hole solutions \cite{gla} have been considered earlier in the semi-classical Einstein's equations with conformal anomaly \cite{Cai:2009ua}, gravity theories with quantum corrections \cite{Cognola:2013fva,Tomozawa:2011gp},  in regularized Lovelock gravity \cite{Casalino:2020kbt}, and also in the $4D$ non-relativistic Horava-Lifshitz theory of gravity \cite{Kehagias:2009is}. 
However, the $4D$ gravity was formulated recently and little is known about the theory, which deserves to be understood better. Nevertheless, recently interesting measures have been taken  to investigate the   $4D$ EGB gravity, including generalizing the  black hole solution to include electric charge in an anti-de Sitter space  \cite{Fernandes:2020rpa}, to the  radiating or nonstatic black hole   solution in Ref. \cite{Ghosh:2020vpc}, which explores some of their properties. The generalization of these static black holes of the   $4D$ EGB gravity  to the axially symmetric or rotating case, Kerr-like, was also addressed \cite{Wei:2020ght,Kumar:2020owy}. In particular, it is  was shown that the rotating black holes solutions for the   $4D$ EGB gravity can be derived starting from exact spherically symmetric spacetime by Newman and Janis \cite{Azreg-Ainou:2014pra}, and they also demonstrated that the    $4D$ EGB gravity is  consistent with the inferred features of M87* black hole shadow. Other probes  in the theory include  studies  of the innermost stable circular orbit (ISCO) \cite{Guo:2020zmf},  its stability and quasi-normal modes \cite{Konoplya:2020bxa},  relativistic star solution \cite{Doneva:2020ped}, noncommutative inspired black holes \cite{Ghosh:2020cob}  thermodynamics geometry \cite{HosseiniMansoori:2020yfj}, regular black holes \cite{Kumar:2020xvu}, derivation of  regularized field equations \cite{Fernandes:2020nbq} and it's  generalisation to Lovelock gravity \cite{Konoplya:2020qqh}.

Here,  we would like to clarify that  the {\it regularization} proposed in  \cite{gla,Cognola:2013fva}, is subject to debate  and many authors raised questions \cite{Ai:2020peo,Hennigar:2020lsl,Shu:2020cjw,Gurses:2020ofy,Mahapatra:2020rds} on it's definiteness.   Many  alternative  {\it regularizations} have also been suggested \cite{Lu:2020iav,Kobayashi:2020wqy,Hennigar:2020lsl,Casalino:2020kbt}.     L\"{u} and Pang  \cite{Lu:2020iav} regularized  EGB gravity by compactifying $D$ dimensional EGB gravity  on $D-4$ dimensional maximally symmetric  space which leads to a  well defined special scalar-tensor theory that belongs to the family of Horndeski  gravity, and is in agreement with the results of \cite{Kobayashi:2020wqy}.  However, the spherically symmetric $4D$  black hole   solution obtained in \cite{gla,Cognola:2013fva} still  remains valid in these regularized theories \cite{Lu:2020iav,Hennigar:2020lsl,Casalino:2020kbt}.  Hence these \textit{regularization} procedures lead to exactly the same  black hole solutions \cite{gla,Cognola:2013fva} at least for the case of $4D$ spherically symmetric spacetimes.  We can confirm that our solution (\ref{sol1})  can be obtained by the  {\it regularization} proposed in Ref. \cite{Hennigar:2020lsl}. Later,  it was demonstrated that linear perturbations are well behaved around maximally symmetric backgrounds \cite{Arrechea:2020evj}, however the equations for second order perturbations are ill-defined \cite{Arrechea:2020evj}.

The main purpose of this paper is to obtain an exact spherically symmetric  black hole  solution, in the   $4D$ EGB gravity, endowed with a clouds of string (CS).  In particular, we explicitly bring out how the effect of a background CS  can modify  black hole solutions and their thermodynamics. We will examine how GB corrections and  background CS alter the qualitative features we know from our experience with black holes in CS, e.g.,  we shall analyse GB corrections on thermodynamic properties of the black holes and also on  local and global stability.  The intense level of activity in string theory has led to the idea that  static Schwarzschild black hole (point mass), may have atmospheres composed of a  CS, which is the one-dimensional analogue  of a cloud of dust. Further, it could  describe a globular cluster with components of dark matter. Strings may have been present in the early universe for the seeding of density inhomogeneities \cite{23,synge}. The  CS for the  Schwarzschild black hole was initiated by by Letelier \cite{Le1}  to show that the event horizon gets enlarged with radius $ r_H={2M}/{(1-a)}$ with $0<a<1$ being the  string cloud parameter \cite{Le1}, thereby enlarging the Schwarzschild radius of the black hole by the factor $(1-a)^{-1}$, and also has consequences on the accretion process on to a black hole \cite{Ganguly:2014cqa}.

\section{Clouds of String  for EGB }
Lovelock demonstrated  that Einstein gravity can be extended by a
 series of higher curvature terms with the  resulting equations of motion still remaining  second order \cite{ll}.
The Lovelock theory is an extension of the  GR to higher dimensions with first and second order terms, respectively, corresponding to the Ricci scalar and a combination of  quadratic curvatures - Gauss-Bonnet.  Action of the 
 Einstein-Gauss-Bonnet (EGB) gravity, which is motivated by the heterotic string theory  \cite{Lanczos:1938sf,Lovelock:1971yv},  by rescaling the GB coupling constant  $\alpha \to \alpha/(D-4)$, yields \cite{gla}
\begin{eqnarray}
\mathcal{I} &=&\frac{1}{2}\int d^{D}x\sqrt{-g}\left[ {\cal R} +\frac{\alpha}{D-4} {\cal L_{GB}} -F_{ab}F^{ab} \right]+\mathcal{I_M},
\label{action1}
\end{eqnarray}
where ${\cal R}$ is the scalar curvature, $\alpha$ is the GB coupling constant  constant, $F_{ab}=\partial _{a}A_{b}-\partial_{b}A_{b}$ is the electromagnetic field tensor, and $A_{b}=-Q/r dt$ is the vector potential.
 $\mathcal{I_M}$ is the action of the matter source which in the present case is a string cloud (\ref{scm}). Varying the action (\ref{action1}), we obtain the equations of motion  \cite{ghosh8,hr}
\begin{eqnarray}
&&{G}_{a b} +\alpha {H}_{a b} =  \mathcal{T}_{ab},
\label{egb3}
\end{eqnarray}
where $G_{ab}$ and $H_{ab}, $ respectively, are the Einstein tensor and the Lanczos tensor:
\begin{eqnarray}
&&G_{ab}=R_{ab}-\frac{1}{2}g_{ab}R,\nonumber\\
&&{H}_{ab}=2\left[RR_{ab}-2R_{a c}R^{c}_{b}-2R^{c d}R_{a c bd} +R_{a}^{~c d e}R_{b c d e}\right]-{1\over 2}g_{ab}{L}_{GB},
\end{eqnarray}
and $\mathcal{T}_{ab} = T_{ab} + T_{ab}^{EM}$. 
We wish to obtain static spherically symmetric black hole solutions of Eq.~(\ref{action1}).  We assume the metric to be of the following form \cite{Ghosh:2020vpc}
\begin{equation}
ds^2 = -f(r)dt^2+\frac{1}{f(r)} dr^2 + r^2 d\Omega_{D-2},
\label{metric}
\end{equation}
where $  d\Omega_{D-2}$ is the metric of a $(D-2)$-dimensional constant curvature space 
and $T_{a b}$ is the energy momentum tensor of matter that we consider as
a cloud of strings. The Nambu-Goto action \cite{Le1} of a string evolving in spacetime is given by
\begin{equation}  
S_{\mathcal{S}} =\int_{\Sigma}  \; m (\gamma)^{-1/2}  d\lambda^{0} d\lambda^{1}= \int_{\Sigma}  \; m \left[-\frac{1}{2} \Sigma^{\mu \nu} \Sigma_{\mu \nu}\right]^{1/2}  d\lambda^{0} d\lambda^{1},
\label{scm}
\end{equation}
where $\gamma $  is the determinant of the $\gamma_{a b}$, the $\gamma_{a b}$ is given by 
\begin{equation}
\gamma_{a b} = g_{\mu \nu} \frac{\partial x^{\mu}}{\partial \lambda^{a}} \frac{\partial x^{\nu}}{\partial \lambda^{b}},
\end{equation}
where $m$ is a positive constant, $\lambda^{0}$ and $\lambda^{1}$ being timelike and  spacelike parameters \cite{synge}. The bivector associated with the string world sheet $\Sigma$ is  given by \cite{Le1}
\begin{equation}
\label{eq:bivector}
\Sigma^{\mu \nu} = \epsilon^{a b} \frac{\partial x^{\mu}}{\partial \lambda^{a}} \frac{\partial x^{\nu}}{\partial \lambda^{b}},
\end{equation}
where $\epsilon^{a b}$ is the Levi-Civita tensor in two dimensions, which is anti-symmetric in $a$ and $b$ given by $\epsilon^{0 1} = - \epsilon^{1 0} = 1$. Further, since $T^{\mu \nu} = 2 \partial \mathcal{L}/\partial g^{\mu \nu}$, 
and finally adapting to parametrization, we get
\begin{equation}
\label{eq:div}
\partial_{\mu} (\sqrt{-\mathbf{g}} \rho \Sigma^{\mu \sigma}) = 0.
\end{equation}
Here  the density $\rho$ and the bivector $\Sigma_{\mu\nu}$ are the functions of $r$ only as we  seek static spherically symmetric solutions. The only surviving component of the bivector $\Sigma$ is $\Sigma^{tr} =  - \Sigma^{rt}$.  Thus, $T^t_t=T^r_r=-\rho \Sigma^{tr}$, and  from Eq.~(\ref{eq:div}), we obtain  $\partial_{r} (\sqrt{r^{D-2} T^t_t}) = 0$, which implies \cite{Le1}
\begin{equation}
T^t_t = T^r_r = \frac{a}{r^{D-2}},
\label{emt2}
\end{equation}
for some real constant $a$.  The stress-energy momentum of CS is same that for the global monopole \cite{mb}.  The monopoles topological defects like cosmic strings and domain walls which were originated during the cooling phase of the early universe \cite{mb,tk}, and they play significant role while investigating the black holes \cite{agk}.

$T_{ab}^{EM}$ is related to the electromagnetic tensor $F_{ab}$ by
\begin{equation}
T^{EM}_{ab} = \frac{1}{4} \left( F_{ac} F_{b}^{c} - \frac{1}{4}
g_{ab} F_{cd} F^{cd} \right) \label{eq:tm},
\end{equation}
which satisfies Maxwell's field equations. 
\subsection{Black hole solution for the   $4D$ EGB}
Many authors generalized the pioneering work of Letelier \cite{Le1}, for instance, in GR \cite{29},  for EGB models \cite{hr}, and in Lovelock gravity \cite{Ghosh:2014pga,Ghosh:2014dqa}.     We are interested in an exact black hole in  the  $4D$ EGB endowed with a CS.  Let us consider the metric (\ref{metric}) with stress tensor and apply the procedure in \cite{gla}.  Now, in the limit $D\to 4$, the $(r,r)$ equation of (\ref{egb3}) reduces to
\begin{equation}
r^5-2r^3\alpha(f(r)-1)f'(r)+r^4(f(r)-1)+r^2\alpha (f(r)-1)^2=a r^4 +\frac {Q^2}{r^4},
\label{eom1}
\end{equation}
which can be integrated to give
\begin{eqnarray}
f_{\pm}(r)=1+\frac{r^2}{2\alpha}\left(1\pm\sqrt{1+4\alpha\left(\frac{2M}{r^3}-\frac{Q^2}{r^4}+\frac{a}{r^2}\right)}\,\right),
\label{sol1}
\end{eqnarray}
by appropriately relating $M$  with integrating constants. Solution
(\ref{sol1}) is an exact solution of the field equation (\ref{egb3}) for stress-energy tensor $ \mathcal{T}_{ab} $
which in absence of CS  and charge $a=Q=0$ reduces to the Glavan and Lin \cite{gla} EGB black hole 
solution, and for $Q=0$ the  charged EGB black hole due to Fernandes \cite{Fernandes:2020rpa}. 
The two  branches of the  solution (\ref{sol1}), in the limit $\alpha \to 0$ or large $r$, behaves asymptotically as 
\begin{eqnarray}
&&f(r)=1-\frac{2M}{r}+\frac{Q^2}{r^2}-a,\nonumber\\
&&f({r})=1+\frac{2M}{r}-\frac{Q^2}{r^2}+a+\frac{r^2}{\alpha}.
\end{eqnarray}
Thus, the $-$ve branch corresponds to the Reissner-Nordstrom solution surrounded by the CS with positive gravitational mass and real charge, whereas the $+$ve branch reduces to the  Reissner-Nordstrom  dS/AdS solution with negative gravitational mass and imaginary charge.  To study the general structure of the solution (\ref{sol1}), we take the limit $r\rightarrow\infty$ or $M=Q=0$ in solution (\ref{sol1}) to obtain
\begin{equation}
\lim_{r\rightarrow \infty} f_+(r)= 1+\frac{r^2}{\alpha},\;\;\; \lim_{r\rightarrow\infty} f_-(r)= 1 +a.
\end{equation}
This means that  the plus ($+$) branch of the solution (\ref{sol1}) is asymptotically de Sitter (anti-de Sitter) depending on the sign of $\alpha$ $(\pm)$, whereas the minus branch of the solution (\ref{sol1}) is asymptotically flat.  
With appropriate choice of the functions $M$ and $Q$, and parameter $\alpha$, one can generate  other solutions. 
However, we shall confine ourself to the $-$ve branch of the solution (\ref{sol1}).   The solution (\ref{sol1}) can be  characterised by the mass $M$,  charge $Q$,  CS parameter $a$ and GB coupling constant  $\alpha$, which is assumed to be positive and for definiteness we call it $4D$ Charged EGB  black hole surrounded by CS.  Interestingly, the semi-classical Einstein's equations with conformal anomaly \cite{Cai:2009ua}, gravity theory with quantum corrections \cite{Cognola:2013fva}, and the regularized Lovelock gravity \cite{Casalino:2020kbt}, have the same form  of the black holes solutions as   $4D$ EGB gravity given by the metric (\ref{metric}) with (\ref{sol1}).  Since the stress-energy tensor of the CS is same as that of the global monopole \cite{mb}, the solution (\ref{sol1}) is also recognized as $4D$ monopole-charged EGB black hole. 
The event horizon is the largest root of  $g^{rr}=0$ of $f(r)=0$, which admits the simple solution 
\begin{equation}\label{horizon}
r_{\pm}=\frac{M\pm\sqrt{M^2-(1-a)(Q^2+\alpha)}}{1-a},
\end{equation}
where $r_{+}$ corresponds to the event horizon while $r_{-}$ is the Cauchy horizon.  Elementary analysis of the zeros of $f(r)=0$ reveals a critical mass \begin{equation}\label{Mc}
M_c= \sqrt{(1-a) (Q^2 + \alpha)},
\end{equation}  such that,
$f(r)=0$ has no zeros if  $M > M_c$, one double zero
if $M = M_c$, and two simple zeros if
$M < M_c$, (Fig. 2). These cases therefore describe, respectively,  $4D$ Charged EGB black hole surrounded by CS with degenerate horizon, and a  non-extreme
black hole with both event and Cauchy horizons.   Clearly the two horizons coincide with the critical radius \begin{equation}\label{Rc}
r_c= \sqrt{\frac{(Q^2+\alpha)}{(1-a)}}.
\end{equation}
\noindent It is clear that the critical value of $M_c$ and $r_c$ depend upon $\alpha$ and $a$.   
\begin{figure*} [h]
	\begin{tabular}{c c c c}
		\includegraphics[width=0.5\linewidth]{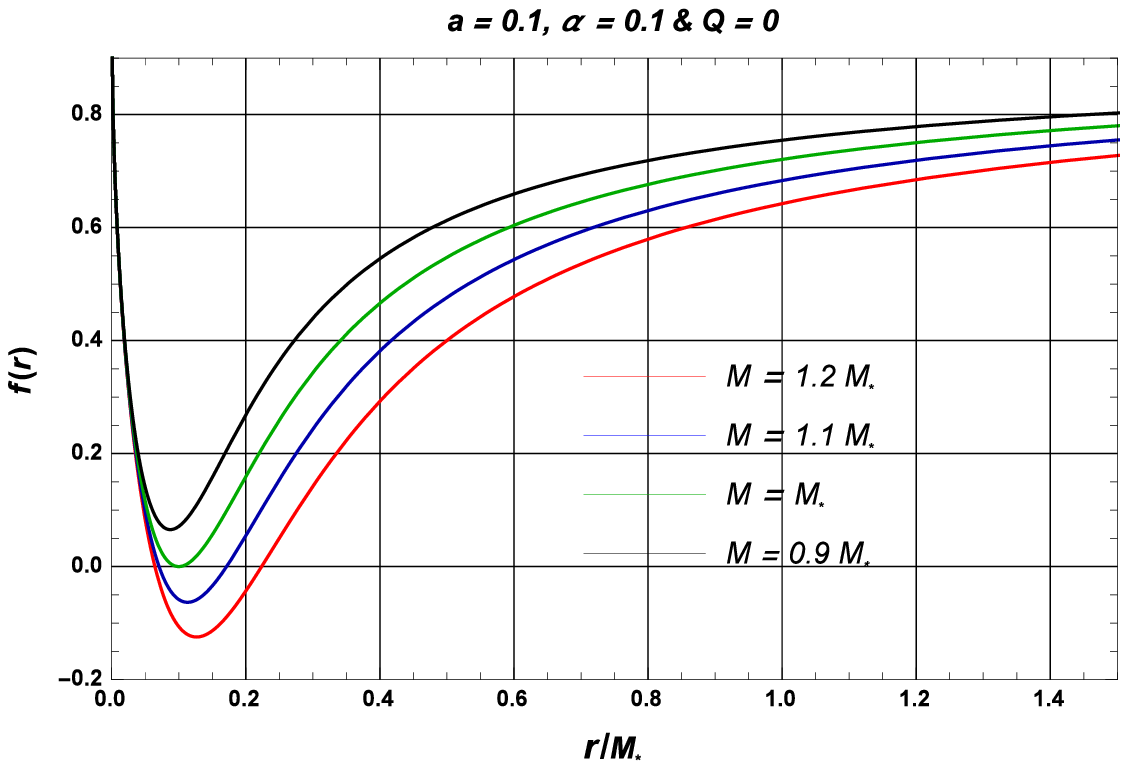}
		\includegraphics[width=0.5\linewidth]{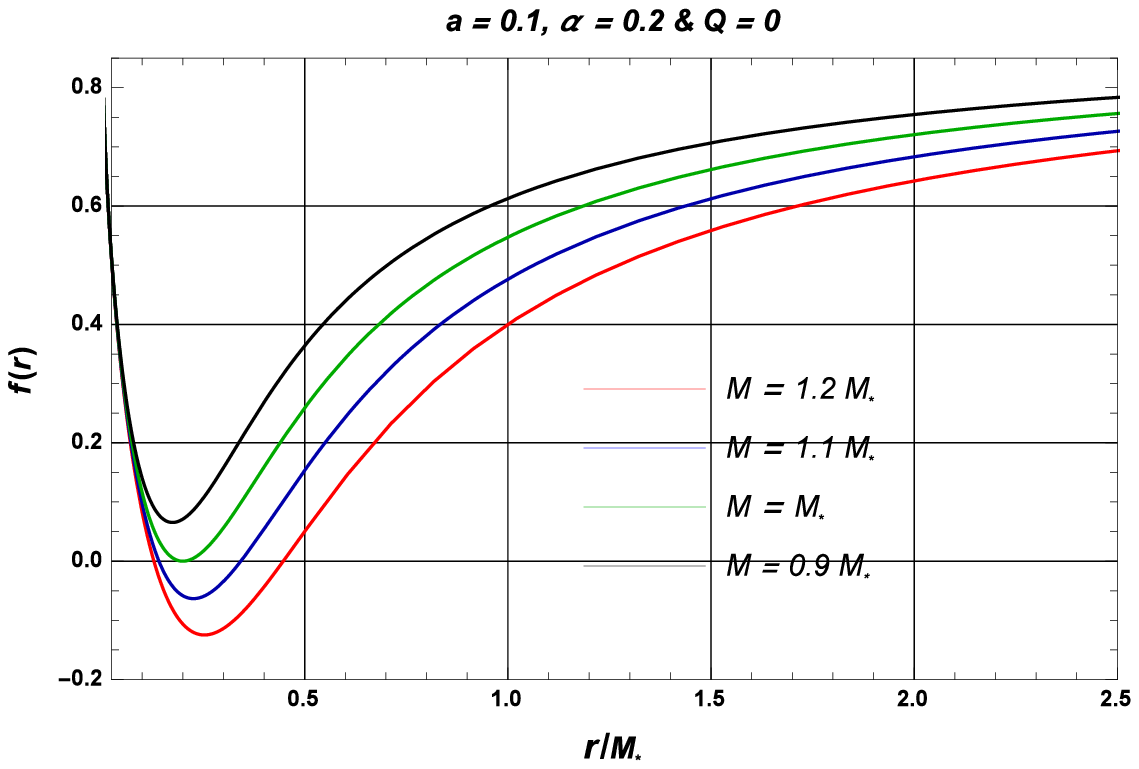}\\
		\includegraphics[width=0.5\linewidth]{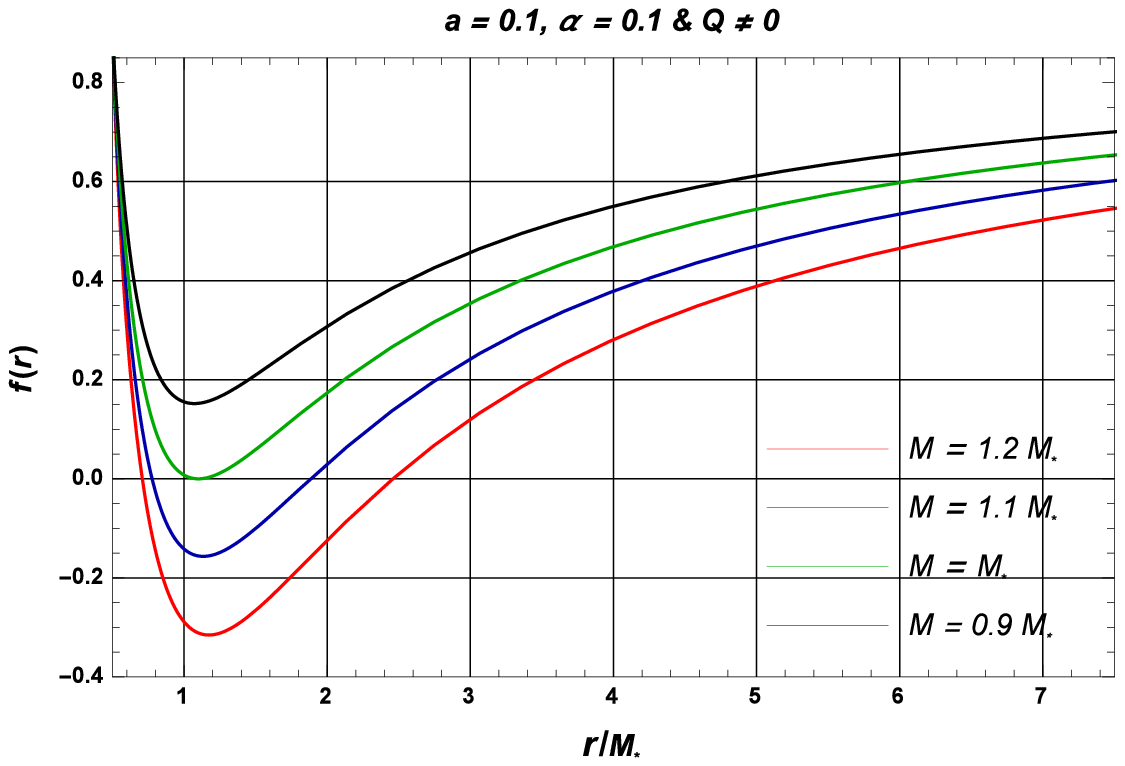}
		\includegraphics[width=0.5\linewidth]{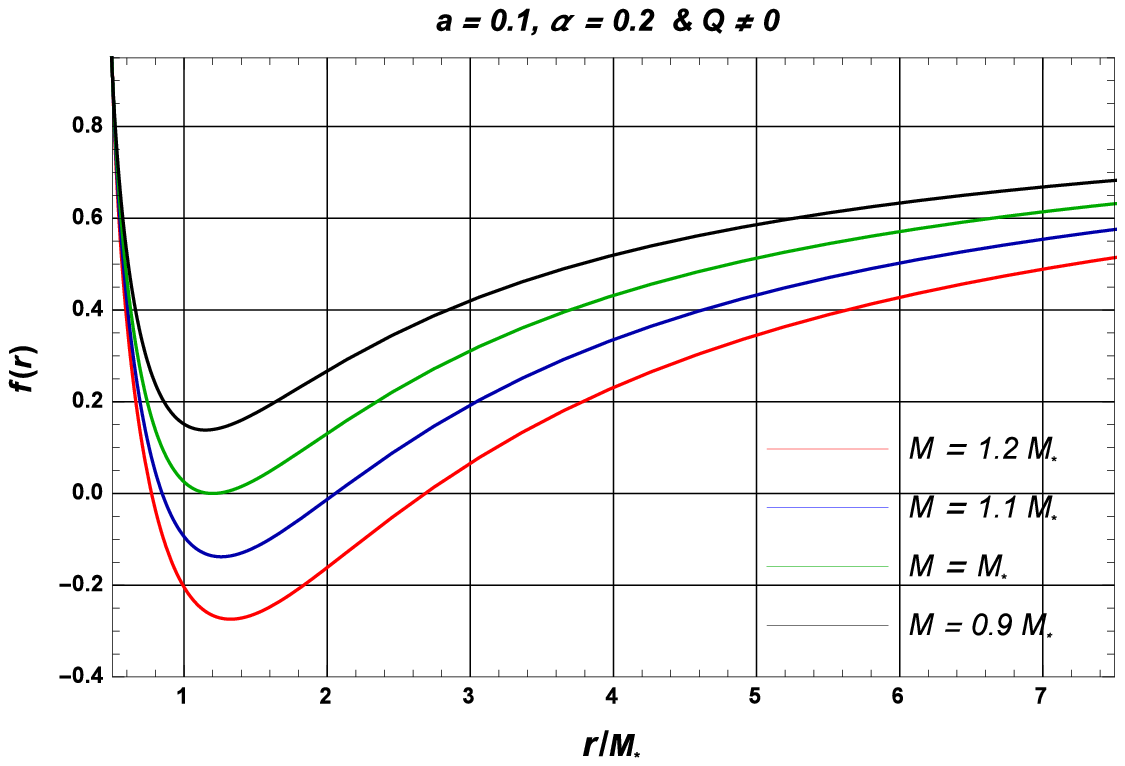}
	\end{tabular}
	\caption{Plot of metric function $f(r)$ vs $r$  for different values of  CS parameter $a$ with GB coupling constant   $\alpha$ = 0.1 and 0.2 for $4D$  EGB black hole surrounded by CS: neutral black hole (upper plots) and 
		charged black hole  (lower plots). }
	\label{f1}
\end{figure*}

\begin{center}
	\begin{table}[h]
		\begin{center}
			\begin{tabular}{l l r l| r l r l r}
				\hline
				\hline
				\multicolumn{1}{c}{ }&\multicolumn{1}{c}{ $\alpha=0.1 $  }&\multicolumn{1}{c}{}&\multicolumn{1}{c|}{ \,\,\,\,\,\, }&\multicolumn{1}{c}{ }&\multicolumn{1}{c}{}&\multicolumn{1}{c}{ $\alpha=0.2 $ }&\multicolumn{1}{c}{}\,\,\,\,\,\,\\
				\hline
				\multicolumn{1}{c}{ \it{M}} & \multicolumn{1}{c}{ $r_-/M_{\star}$ } & \multicolumn{1}{c}{ $r_+/M_{\star}$ }& \multicolumn{1}{c|}{$\delta$}&\multicolumn{1}{c}{\it{ M}}& \multicolumn{1}{c}{$r_-/M_{\star}$} &\multicolumn{1}{c}{$r_+/M_{\star}$} &\multicolumn{1}{c}{$\delta$}   \\
				\hline
				\multicolumn{1}{c}{ \bf{$Q=0 \,\,\&\,\,  a=0.1$}} & \multicolumn{1}{c}{  } & \multicolumn{1}{c}{ }& \multicolumn{1}{c|}{}&\multicolumn{1}{c}{\it{ }}& \multicolumn{1}{c}{} &\multicolumn{1}{c}{} &\multicolumn{1}{c}{}   \\
				\hline
				\,\,\, 1.2 $M_{\star}$\,\,& \,\,0.065\,\, &\,\,  0.224\,\,& \,\,0.159\,\,&1.2 $M_{\star}$&\,\, 0.128\,\,&\,\,0.448\,\,&\,\,0.320\,\,
				\\
				\
				\,\, 1.1 $M_{\star}$\,\, & \,\,0.069\,\, &\,\, 0.171\,\,& \,\,0.102\,\,&1.1 $M_{\star}$&\,\,0.138\,\,&\,\,0.334\,\,&\,\,0.196\,\,
				\\
				$ \,\,\,M_{\star}=0.3$\,\, &  \,\,0.11\,\,  &\,\,0.11\,\,&\,\,0\,\,&$ M_{\star}=0.4242 $&\,\,0.22\,\,&\,\,0.22\,\,&\,\,0\,\,
				\\
				\multicolumn{1}{c}{ \bf{$Q\neq0 \,\,\&\,\,  a=0$}} & \multicolumn{1}{c}{  } & \multicolumn{1}{c}{ }& \multicolumn{1}{c|}{}&\multicolumn{1}{c}{\it{ }}& \multicolumn{1}{c}{} &\multicolumn{1}{c}{} &\multicolumn{1}{c}{}   \\
				\hline
				\,\,\, 1.2 $M_{\star}$\,\,& \,\,0.562\,\, &\,\,  1.954\,\,& \,\,1.392\,\,&1.2 $M_{\star}$&\,\, 0.587\,\,&\,\,2.041\,\,&\,\,1.454\,\,
				\\
				\
				\,\, 1.1 $M_{\star}$\,\, & \,\,0.673\,\, &\,\, 1.634\,\,& \,\,0.961\,\,&1.1 $M_{\star}$&\,\,0.702\,\,&\,\,1.707,\,&\,\,1.01\,\,
				\\
				$\,\,\, M_{\star}=1.048$\,\, &  \,\,1.04\,\,  &\,\,1.04\,\,&\,\,0\,\,&$ M_{\star}=1.09$&\,\,1.09\,\,&\,\,1.09\,\,&\,\,0\,\,
				\\
				\hline
				\multicolumn{1}{c}{ \bf{$Q\neq0 \,\,\&\,\,  a=0.1$}} & \multicolumn{1}{c}{  } & \multicolumn{1}{c}{ }& \multicolumn{1}{c|}{}&\multicolumn{1}{c}{\it{ }}& \multicolumn{1}{c}{} &\multicolumn{1}{c}{} &\multicolumn{1}{c}{}   \\
				\hline
				\,\,\, 1.2 $M_{\star}$\,\,& \,\,0.593\,\, &\,\,  2.059\,\,& \,\,1.46\,\,&1.2 $M_{\star}$&\,\, 0.626\,\,&\,\,2.70\,\,&\,\,2.07\,\,
				\\
				\
				\,\, 1.1 $M_{\star}$\,\, & \,\,0.709\,\, &\,\, 1.722
				\,\,& \,\,1.01\,\,&1.1 $M_{\star}$&\,\,0.750\,\,&\,\,2.06\,\,&\,\,1.31\,\,
				\\
				$\,\,\, M_{\star}=0.994$\,\, &  \,\,1.1\,\,  &\,\,1.1\,\,&\,\,0\,\,&$ M_{\star}=1.039$&\,\,1.20\,\,&\,\,1.20\,\,&\,\,0\,\,
				\\
				\hline
				\multicolumn{1}{c}{ \bf{$Q\neq0 \,\,\&\,\,  a=0.3$}} & \multicolumn{1}{c}{  } & \multicolumn{1}{c}{ }& \multicolumn{1}{c|}{}&\multicolumn{1}{c}{\it{ }}& \multicolumn{1}{c}{} &\multicolumn{1}{c}{} &\multicolumn{1}{c}{}   \\
				\hline
				\,\,\, 1.2 $M_{\star}$\,\,& \,\,0.672\,\, &\,\,  2.335\,\,& \,\,1.663\,\,&1.2 $M_{\star}$&\,\, 0.702\,\,&\,\,2.43\,\,&\,\,1.728\,\,
				\\
				\
				\,\, 1.1 $M_{\star}$\,\, & \,\,0.804\,\, &\,\, 1.149\,\,& \,\,0.345\,\,&1.1 $M_{\star}$&\,\,0.840\,\,&\,\,2.041\,\,&\,\,1.21\,\,
				\\
				$\,\,\, M_{\star}=0.877$\,\, &  \,\,1.25\,\,  &\,\,1.25\,\,&\,\,0\,\,&$ M_{\star}=0.916$&\,\,1.309\,\,&\,\,1.309\,\,&\,\,0\,\,
				\\
				\hline
				\hline
				\hline
			\end{tabular}
		\end{center}
		\caption{Cauchy ($r_{-}$) and event ($r_{+}$) horizons, and $\delta=r_+-r_-$ for  the $4D$  EGB black hole surrounded by CS.}
		\label{tab:temp1}
	\end{table}
\end{center}
\section{Black hole thermodynamics}
Having completed the discussion of our black hole solutions, we shall discuss  the thermodynamical properties of, for both charged and neutral,  the $4D$  EGB  black hole surrounded by CS  treating $a$ as an external parameter. Hawking and Page demonstrated \cite {hp} that asymptotically anti-de Sitter Schwar--zschild black holes are thermally favoured when their temperature is sufficiently high. The phase transition has been widely studied for others asymptotically AdS black holes in the context of EGB gravity \cite{ipn,noc,Anninos:2008sj}.  Now, we are going to study   thermodynamic quantities associated with the  $4D$  EGB black hole surrounded by CS.  We note that the gravitational mass of a black hole is determined by $f(r_+)=0$ \cite{Ghosh1:2018bxg}, which reads 
\begin{eqnarray}
M_+= \frac{r_+}{2} \left( 1 + \frac{Q^2+ \alpha}{r_+^{2}}-a \right).
\label{mass1}
\end{eqnarray}
Eq. (\ref{mass1}) reduces to the  black hole mass 
 \[ M_+= \frac{r_+}{2} \left( 1 + \frac{Q^2+ \alpha}{r_+^{2}} \right), \] 
 for the  $4D$ Charged EGB black hole \cite{Fernandes:2020rpa} when $a=0$,  to the  $4 D$  Reissner-Nordstrom  black hole for $\alpha\rightarrow 0, a=0$, and  in the absence of charge $(Q=0)$ it  reduces to the mass of $4D$ EGB black hole surrounded by CS, and we obtain mass for the Schwarzschild black hole surrounded by CS as $M_{+} =((1-a) r_{+})/2$ \cite{Ghosh:2014pga}.  It is evident from the Eq.~(\ref{Mc}) that black hole enrich with CS  has higher critical mass and so is  the event horizon radius  with increase in the CS parameter ($a$) and increases with the  Gauss-Bonnet coupling constant  $\alpha$ (cf. Fig. \ref{f1} and Table \ref{f1}). 

A black hole at high (low) temperature is stable (unstable), and there can occur a phase transition between the thermal AdS and the AdS black hole  at some critical temperature \cite{hp}. 
The Hawking temperature of the black hole is defined  to be proportional  to the surface gravity  $\kappa$ by $T=\kappa/2\pi$, where $\kappa$  is given by
\begin{equation} 
\kappa=\frac{1}{2\pi}\left(-\frac{1}{2}\nabla_{\mu}\xi_{\nu}\nabla^{\mu}\xi^{\nu}\right)^{1/2},
\end{equation}
and $\xi^{\mu}=\partial/\partial t$ is a Killing vector.  The black hole has a  Hawking temperature defined by $T=\kappa/2\pi$ \cite{Ghosh1:2018bxg}, where $\kappa$ is the surface gravity given. Applying this formula to our metric function (\ref{sol1}), the Hawking temperature for the $4D$ EGB black hole with CS  reads
\begin{eqnarray}
T_+=\frac{1}{4\pi r_+}\left(\frac{r_+^2-ar_+^2-(Q^2+\alpha)}{r_+^2+2\alpha}\right).
\label{temp1}
\end{eqnarray}
The temperature of $4D$ charged  EGB  black hole with CS (\ref{temp1}) reduces to the temperature of $4D$ charged EGB black hole \cite{Fernandes:2020rpa} when $a=0$, $4D$  EGB black hole \cite{gla} in the limit of $a=0,\,Q=0$,  $4D$  black hole with CS parameter  when $Q=0 $ and $\alpha \to 0$ \cite{Ghosh:2014pga}, and  also to  Schwarzschild black hole surrounded by CS: $T_+=(1-a)/4\pi r_+$  when $Q=0$ and $\alpha \to 0$ \cite{Ghosh:2014pga} .

\begin{center}
	\begin{table}[h]
		\begin{center}
			\begin{tabular}{l|l r l r l| r l r r r}
				\hline
				\hline
				\multicolumn{1}{c}{ }&\multicolumn{1}{c}{ }&\multicolumn{1}{c}{ }&\multicolumn{1}{c}{$\alpha=0.1 M^2$  }&\multicolumn{1}{c}{ \,\,\,\,\,\, }&\multicolumn{1}{c|}{ }&\multicolumn{1}{c}{  }&\multicolumn{1}{c}{ }&\multicolumn{1}{c}{$\alpha=0.2 M^2$}\,\,\,\,\,\,\\
				\hline
				\multicolumn{1}{c}{ }&\multicolumn{1}{c}{ $Q/M=0$}&\multicolumn{1}{c}{ }&\multicolumn{1}{c}{  }&\multicolumn{1}{c}{ \,\,\,\,\,\, }&\multicolumn{1}{c}{ }&\multicolumn{1}{c}{  }&\multicolumn{1}{c}{ }&\multicolumn{1}{c}{}\,\,\,\,\,\,\\
				\hline
				\multicolumn{1}{c|}{a } &\multicolumn{1}{c}{ 0 } &\multicolumn{1}{c}{ 0.10 } & \multicolumn{1}{c}{ 0.375 }& \multicolumn{1}{c}{0.45}& \multicolumn{1}{c|}{0.55} &\multicolumn{1}{c}{ 0}&\multicolumn{1}{c}{0.10} &\multicolumn{1}{c}{ 0.375}   & \multicolumn{1}{c}{0.45}& \multicolumn{1}{c}{0.55} \\
				\hline
				$r_c^T/M$ & 0.724& 0.754 &0.834 &0.885 &0.944&1.02&1.04&1.135&1.186&1.303
				\\
				$T_+^{Max}M$&0.064& 0.056 &0.035& 0.0302 & 0.0231&0.0457&0.0397& 0.0253&0.0213&0.0164\\
				\hline
				\multicolumn{1}{c}{ }&\multicolumn{1}{c}{ $Q=0.70M$}&\multicolumn{1}{c}{ }&\multicolumn{1}{c}{  }&\multicolumn{1}{c}{ \,\,\,\,\,\, }&\multicolumn{1}{c}{ }&\multicolumn{1}{c}{  }&\multicolumn{1}{c}{ }&\multicolumn{1}{c}{}\,\,\,\,\,\,\\
				\hline
				\multicolumn{1}{c|}{a } &\multicolumn{1}{c}{ 0 } &\multicolumn{1}{c}{ 0.10 } & \multicolumn{1}{c}{ 0.375 }& \multicolumn{1}{c}{0.45}& \multicolumn{1}{c|}{0.55} &\multicolumn{1}{c}{ 0}&\multicolumn{1}{c}{0.10} &\multicolumn{1}{c}{ 0.375}   & \multicolumn{1}{c}{0.45}& \multicolumn{1}{c}{0.55} \\
				\hline
				$r_c^T/M$ &1.222& 1.462 &1.503 &1.512 &1.521&1.311&1.554&1.563&1.602&1.628
				\\
				$T_+^{Max}M$&0.0431& 0.0311 &0.0302& 0.0286 & 0.0277&0.0341&0.0269& 0.0263&0.0251&0.0242\\
				\hline
				\multicolumn{1}{c}{ }&\multicolumn{1}{c}{ $a=0.15$}&\multicolumn{1}{c}{ }&\multicolumn{1}{c}{  }&\multicolumn{1}{c}{ \,\,\,\,\,\, }&\multicolumn{1}{c}{ }&\multicolumn{1}{c}{  }&\multicolumn{1}{c}{ }&\multicolumn{1}{c}{}\,\,\,\,\,\,\\
				\hline
				\multicolumn{1}{c|}{Q/M } &\multicolumn{1}{c}{ 0 } &\multicolumn{1}{c}{ 0.50 } & \multicolumn{1}{c}{ 0.55 }& \multicolumn{1}{c}{0.60}& \multicolumn{1}{c|}{0.65} &\multicolumn{1}{c}{ 0}&\multicolumn{1}{c}{0.50} &\multicolumn{1}{c}{ 0.55}   & \multicolumn{1}{c}{0.60}& \multicolumn{1}{c}{0.65} \\
				\hline
				$r_c^T/M$ & 0.7627&1.207 &1.285 &1.354 &1.44&1.084&1.428&1.498&1.548&1.628
				\\
				$T_+^{Max}M$&0.0527& 0.0353 &0.0335& 0.0317 & 0.0302&0.0372&0.0293& 0.0282&0.0272&0.0261\\
				\hline
				\hline
			\end{tabular}
		\end{center}
		\caption{The maximum Hawking temperature ($T_+^{max}$) at critical radius ($r_c^{T}$) for the $4D$  EGB black hole surrounded by CS. $T_+^{max}$ decreases when either of $a$, $Q$ and $\alpha$ increases.}
		\label{tab:temp}
	\end{table}
\end{center}

\begin{figure*} [h]
	\begin{tabular}{c c c c}
		\includegraphics[width=0.5\linewidth]{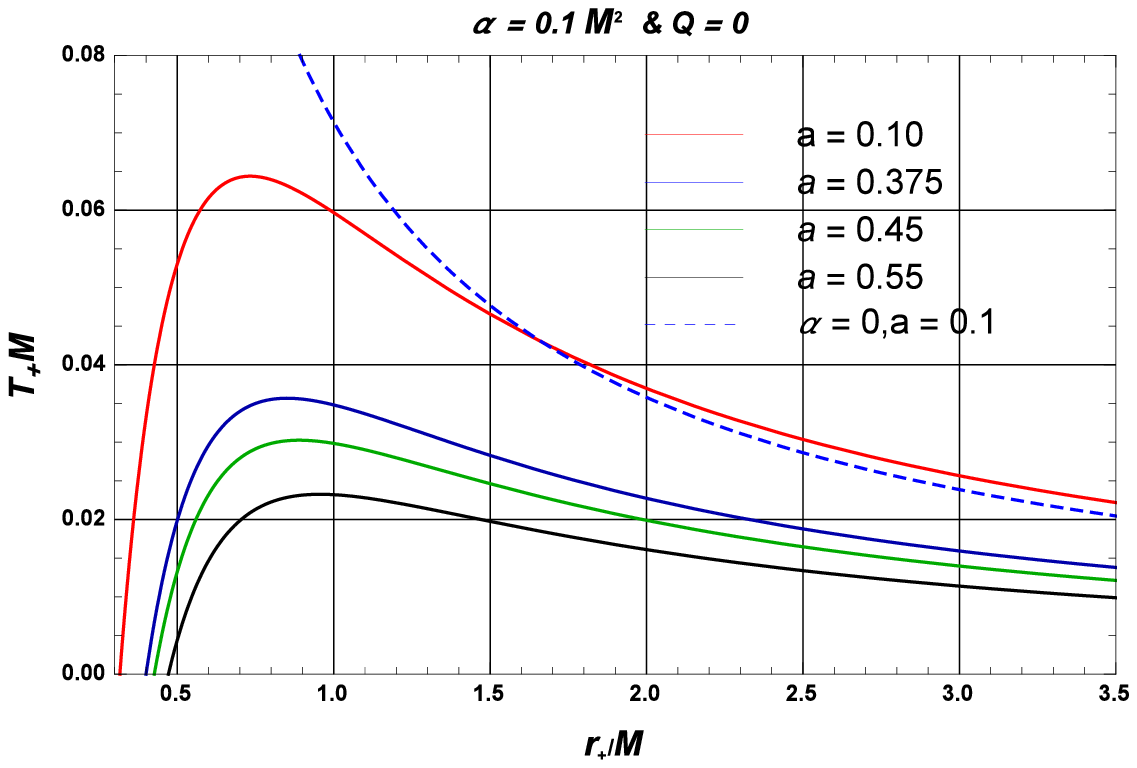}
		\includegraphics[width=0.5\linewidth]{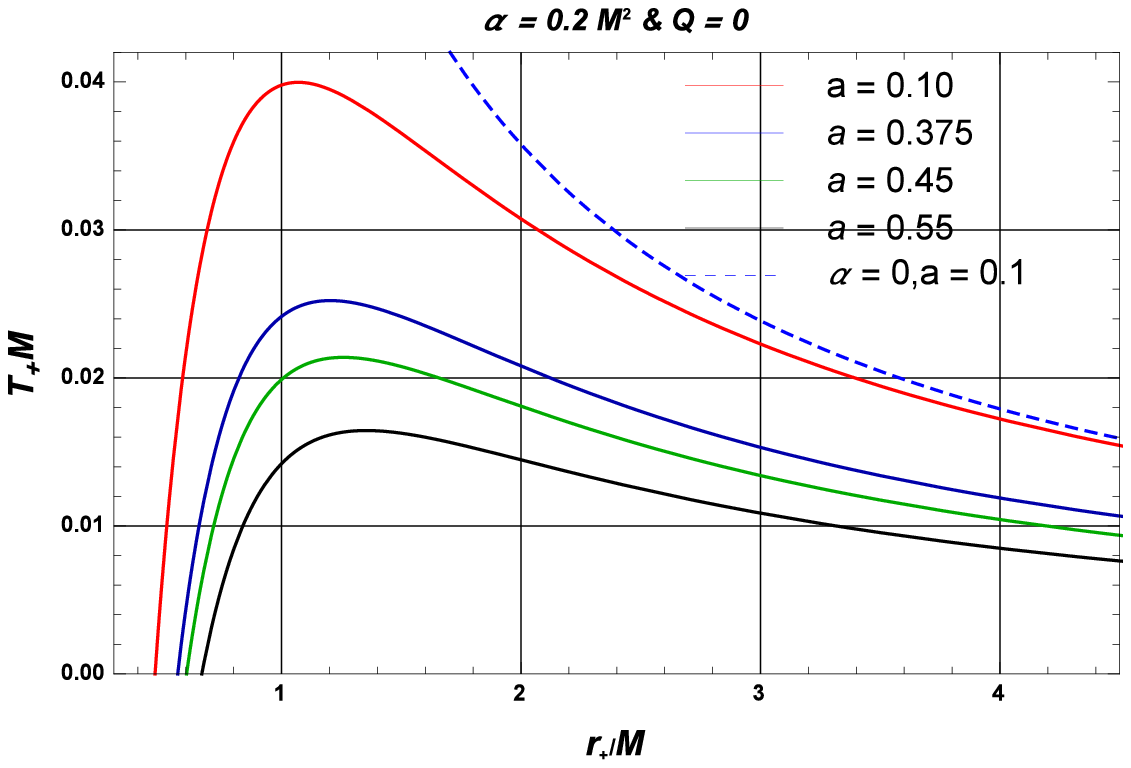}\\
\includegraphics[width=0.5\linewidth]{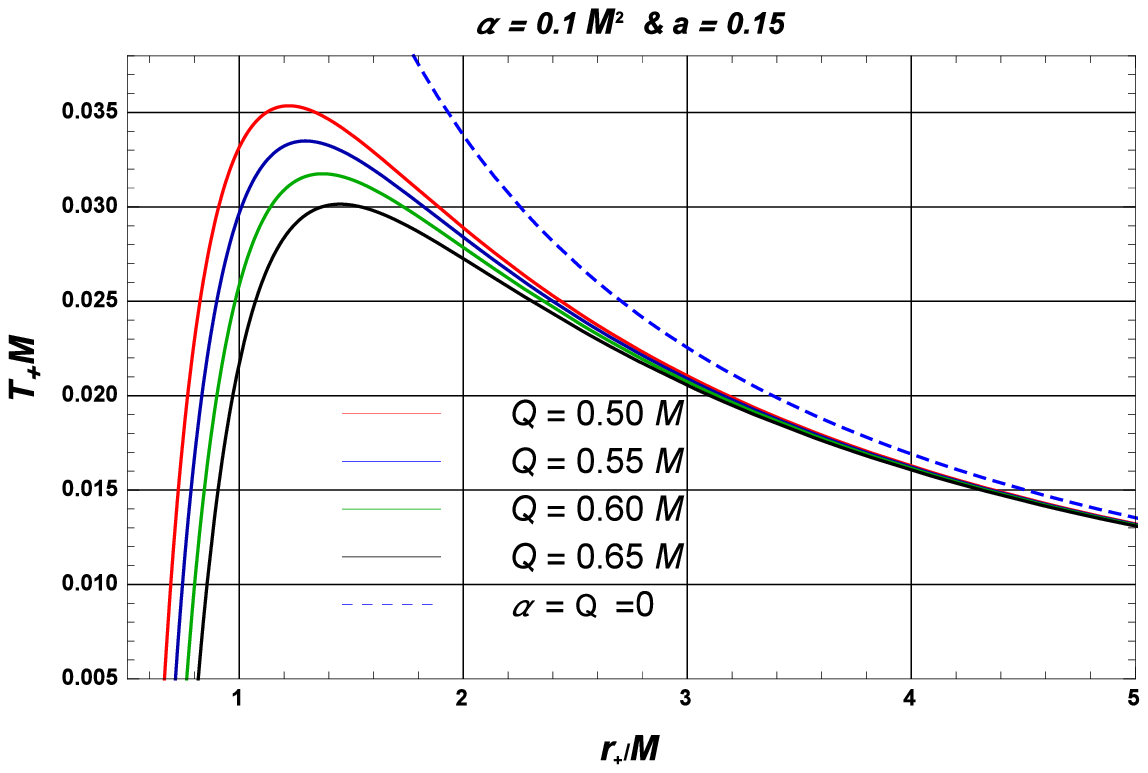}
\includegraphics[width=0.5\linewidth]{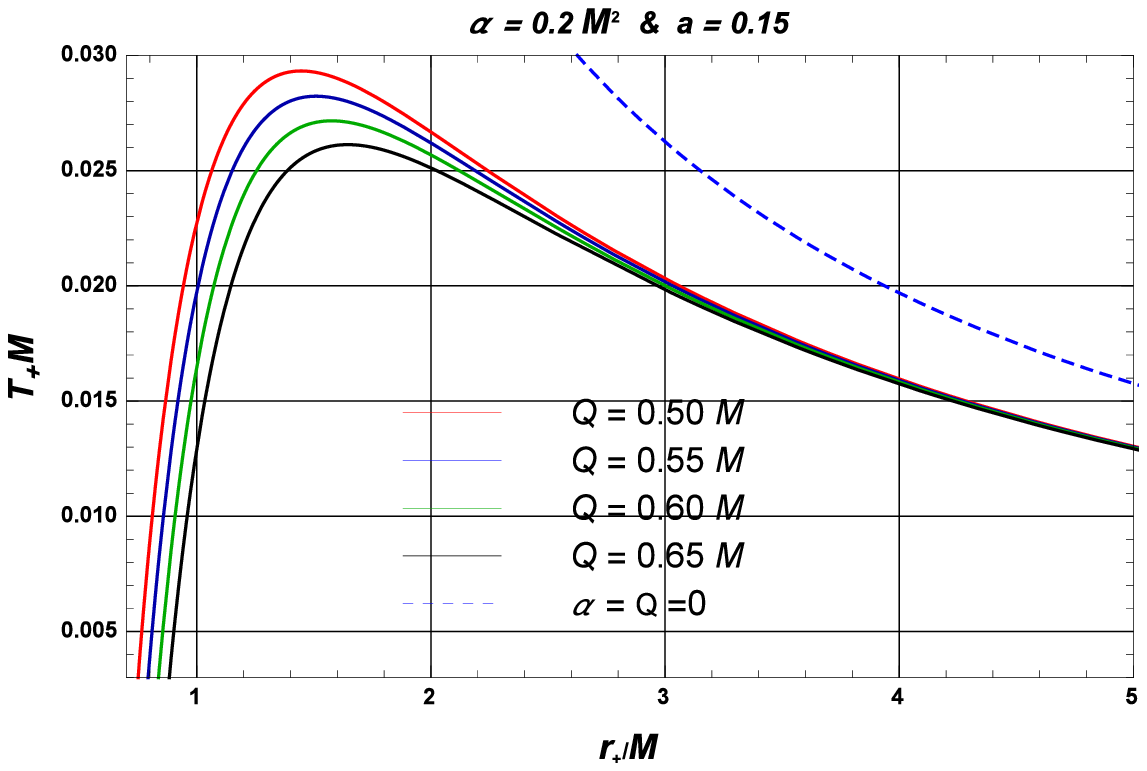}
	\end{tabular}
	\caption{Plot of temperature $T_+$  vs horizon radius $r$ for both neutral (upper) and charged (lower) $4D$  EGB black hole surrounded by CS for different values of GB coupling constant  $\alpha$. $T_+^{max}$ decreases decreases when either of $a$, $Q$ and $\alpha$ increases.}
	\label{fig:th1} 
\end{figure*}

In  Fig. \ref{fig:th1}, we have  shown the Hawking  temperature of the $4D$ charged EGB surrounded by $CS$ grows to a maximum $T^{max}_+$   then drops to zero temperature.  A maximum of the  Hawking temperature occurs at the critical radius shown in Table \ref{tab:temp}. It turns out that the maximum value of the Hawking temperature decreases with increase in the values of the  CS parameter $a$ and GB coupling constant  $\alpha$ for both charged and uncharged black holes  (cf. Fig. \ref{fig:th1} and Table \ref{tab:temp}). 

We  calculate another useful quantity associated with the black hole, in terms  of  horizon radius $r_+$, known as entropy.  The black hole behaves as a thermodynamic system; quantities associated with it must obey the first law of thermodynamics
\begin{eqnarray}\label{1law}
dM_+ = T_+dS_+ +\phi\, dQ,
\end{eqnarray}
where $\phi$ is the potential of the black hole and $dM=TdS$. Hence the entropy \cite{thermo,Ghosh1:2018bxg} can be obtained by  integrating Eq. (\ref{1law}), with $Q$ = constant, as
\begin{eqnarray}
S_+=\int\frac{1}{ T_+}\frac{\partial M_+}{\partial r_+}dr_+= \pi r_{+}^2+ 2\pi \alpha \log(r_+^2)+ S_0 = \frac{A}{4}+2\pi \alpha \log\left(\frac{A}{A_0}\right), 
\label{entropy1}
\end{eqnarray}
with $A= 4 \pi r_+^2$ and $ A_{0}$ is constant having the unit of area. 
This is the standard  area law with logarithmic corrections known as the Bekenstein-Hawking area law \cite{Cai:2009ua}. It is interesting to note that the entropy (\ref{entropy1}) is independent of the  string cloud background and charge  $Q$ \cite{Ghosh1:2018bxg}.   

\subsection{Global stability}
In order to obtain more detail of the thermodynamical equilibrium of the $4D$ EGB black hole, we are interested to study the behaviour of the Gibbs free energy \cite{Ghosh1:2018bxg,Anninos:2008sj,ipn}.  We are essentially going to search, as in the usual Hawking-Page transition, for regions  where the free energy is negative, and identify these regions with black holes that are thermally favored over the reference background \cite{noc,Anninos:2008sj}. 
We turn to calculate the free energy by using $(F_+ =M_+-T_+S_+)$ \cite{Ghosh1:2018bxg}. Upon using the $M_+, T_+$ and $S_+$, the free energy is given by

\begin{equation}
F_+= \frac{1}{4r_+} \left[ 2 \left(Q^2 + (1-a)r_{+}^2 + \alpha \right) + \frac{\left(Q^2 - (1-a)r_{+}^2 + \alpha \right) (r_{+}^2 + 2 \alpha \log(\frac{A}{A_0}) ) }{r_{+}^2+2 \alpha}\right]
\label{freef}
\end{equation}
\begin{figure*} [h]
	\begin{tabular}{c c c c}
		\includegraphics[width=0.5\linewidth]{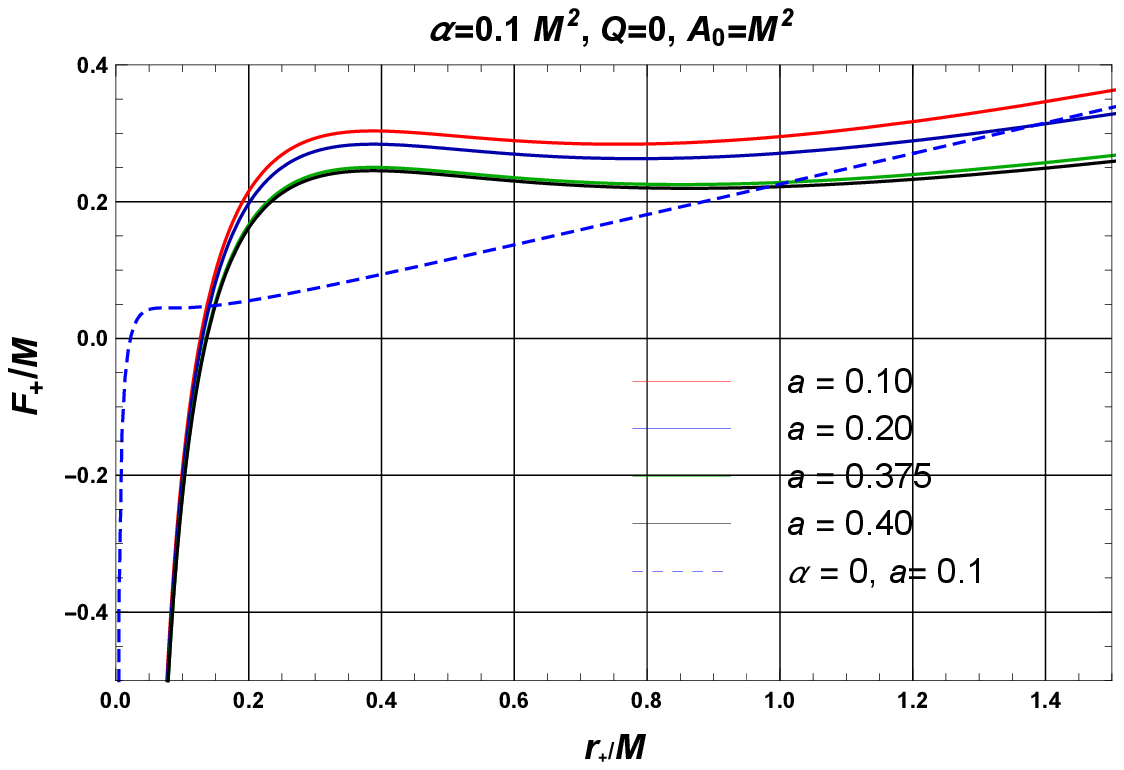}
		\includegraphics[width=0.5\linewidth]{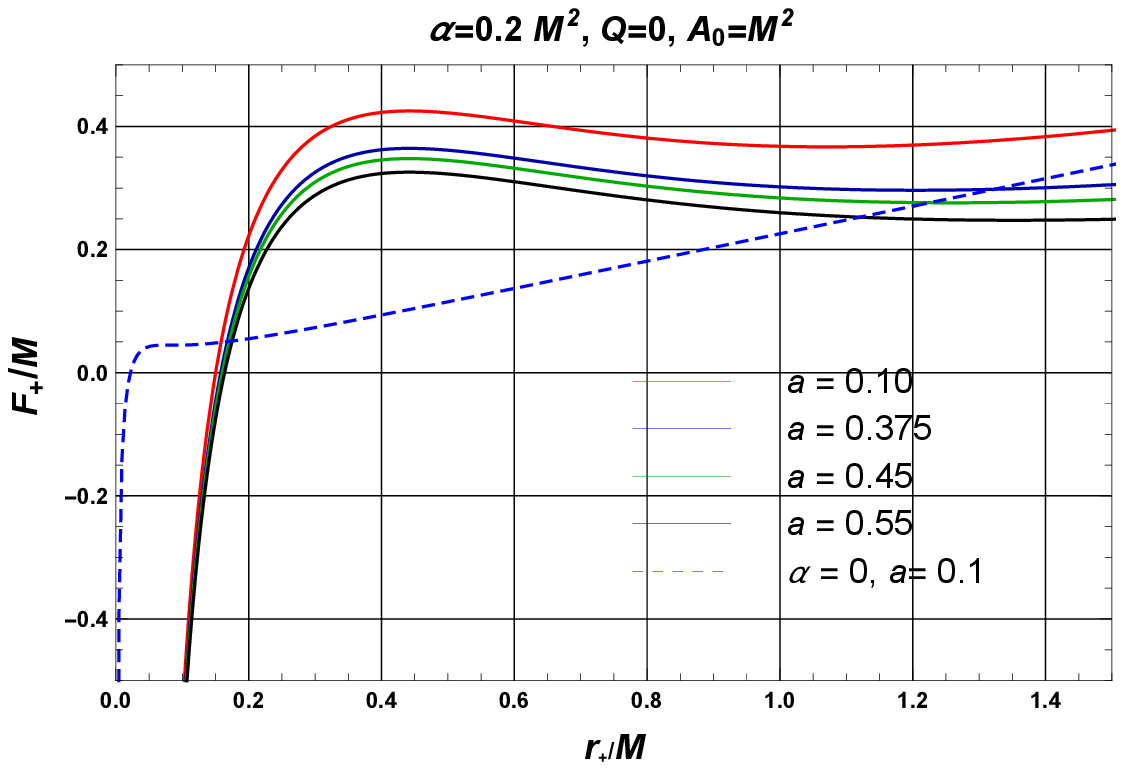}\\
\includegraphics[width=0.5\linewidth]{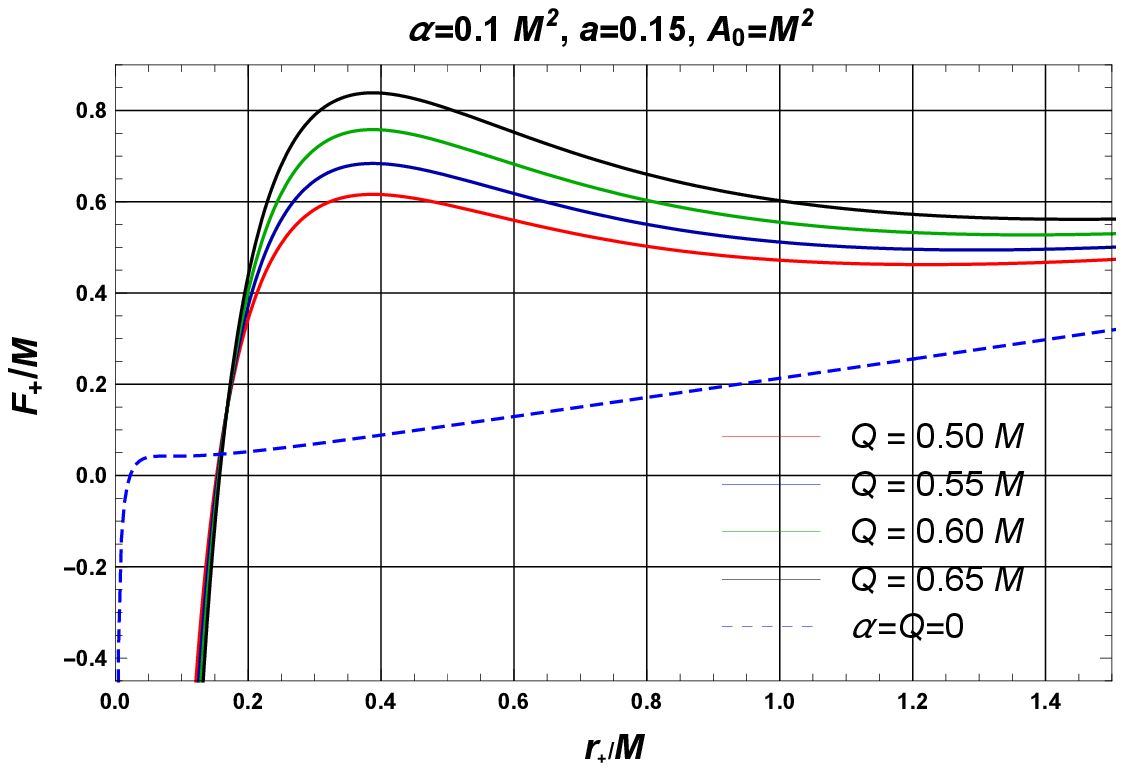}
\includegraphics[width=0.5\linewidth]{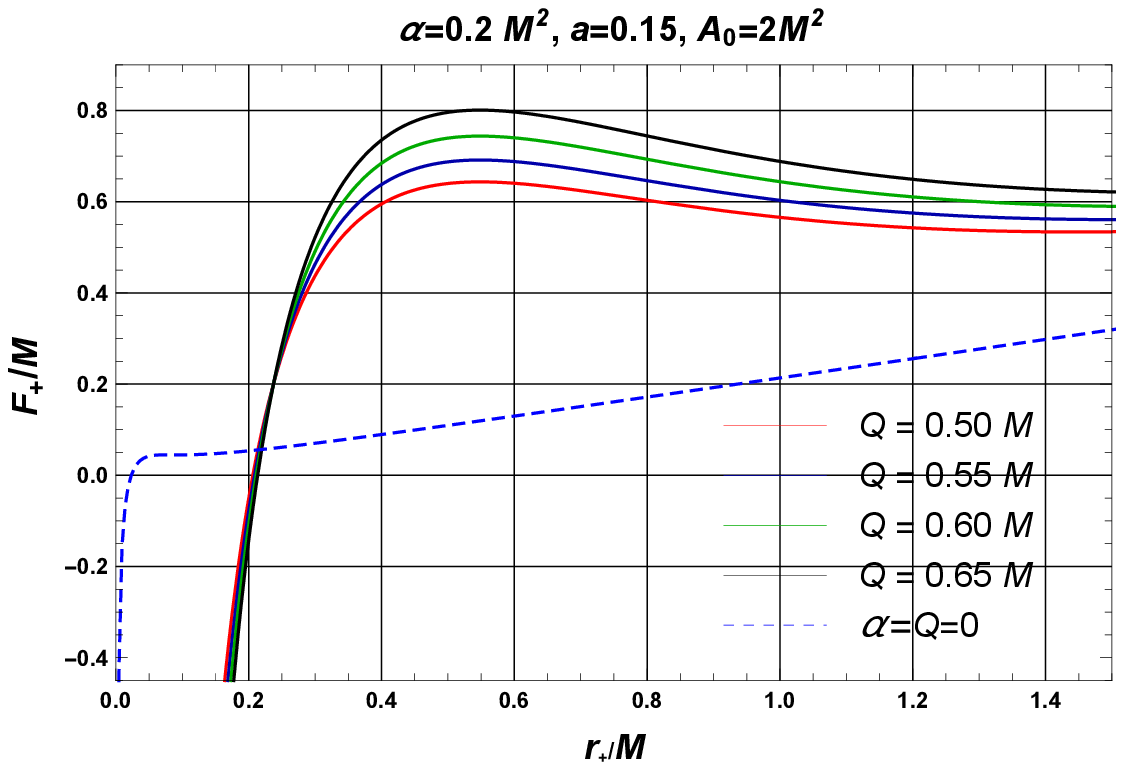}
	\end{tabular}
	\caption{ Plot of free energy $F_+$  vs horizon radius $r_+$ for both neutral (upper) and charged (lower) $4D$  EGB black hole surrounded by CS  for different values of GB coupling constant  $\alpha$.}
	\label{free1}
\end{figure*}
\noindent The Gibbs free energy for various values  of CS parameter $a$ is depicted  in the Fig. \ref{free1} which suggests that it is positive for larger $r_+$.  By analyzing the  free energy it can be revealed  that it  is positive for the larger values of the $r_+$, so the black hole, in the thermodynamical sense, is globally stable the small black hole jumps into a large black hole via a first order phase transition.  Whereas for smaller values of $r_+$ it is negative \cite{hr,Anninos:2008sj}, i.e., the $4D$ EGB black hole  globally stable black hole.   It is very interesting to note that surrounded  by a cloud of strings, the Hawking–Page phase \cite{hp} transition is not possible and rather  we have phase transition from globally thermodynamically stable small black holes with negative heat capacity $F_{+}<0$ to  unstable  large black holes.  In addition, we can find the critical temperature $T_c $,  in terms of the horizon radius $r_+$, by solving $F(T_c)=0$, which result in 
\begin{equation}
T_{c}=\frac{Q^2+r_+^2(1-a)+\alpha}{2\pi r_+(r_+^2+4\alpha\log(r_+))}
\end{equation}
This Hawking-Page is marked by the point where $T=T_c$, whereas  for $T>T_c$ we find that the black hole solution is thermally favoured globally with respect to the reference background solution. While for $T<T_c$ the reference background solution is globally favored \cite{Anninos:2008sj,hr}.  
\subsection{Local stability}
Next we proceed to analyse  the local thermodynamic structure by computing the specific heat which help understand us about the thermal stability of the black hole under temperature fluctuations \cite{Anninos:2008sj,hr}.  It turns out that a black hole configuration can be  global thermodynamic preference to the reference background, but it is  possible that the black hole can still be locally unstable.   We analyse how the background CS affects  the thermodynamic stability of the $4D$ charged EGB black hole  by investigating the  heat capacity $C_+$. The stability of the black hole is related to sign of the heat capacity $C_+$. When $C_+>0$ the black hole is stable and $C_+<0$ means it's unstable. The heat capacity of the black hole is given \cite{cai}
\begin{eqnarray}\label{SH}
C_+&=&\frac{\partial{M_+}}{\partial{T_+}}=\left(\frac{\partial{M_+}}{\partial{r_+}}\right)\left(\frac{\partial{r_+}}{\partial{T_+}}\right).
\label{sp1}
\end{eqnarray}
Substituting the values of mass and temperature from  Eqs. (\ref{mass1}) and (\ref{temp1}) in Eq. (\ref{sp1}) , we obtain the  heat capacity of  the $4D$ charged EGB surrounded by CS as
\begin{equation}
C_+=-\frac{ 2\pi r_+^2 (r_+^2+2\alpha)^2\left(\frac{Q^2}{r_+^2}+\frac{\alpha}{r_+^2}-(1-a)\right)}{(5-2\alpha)r_+^2\alpha+2\alpha^2+Q^2(3r_+^2+2\alpha)-(1-a)r_+^4},
\label{28}
\end{equation}
The heat capacity (\ref{28}), depends on the Gauss-Bonnet coefficient $\alpha$, a string cloud parameter $a$, and the charge $q$ and, in the limit $\alpha\rightarrow0$, one regains the analogous GR case, i.e., the Eq. (\ref{28}) becomes 
\begin{equation}
C_+=\frac{- 2\pi r_+^2 \left[Q^2-(1-a)r_+^2\right]}{3Q^2-(1-a)r_+^2}.
\label{26}
\end{equation}
The heat capacity (\ref{28}), in the absence of charge ($Q=0$), reduces for the $4D$ EGB surrounded by CS
\begin{equation}
C_+=-\frac{ 2\pi r_+^2 (r_+^2+2\alpha)^2\left(\frac{\alpha}{r_+^2}-(1-a)\right)}{(5-2\alpha)r_+^2\alpha+2\alpha^2-(1-a)r_+^4},
\label{27}
\end{equation}
\begin{figure*} [h]
	\begin{tabular}{c c c c}
		\includegraphics[width=0.5\linewidth]{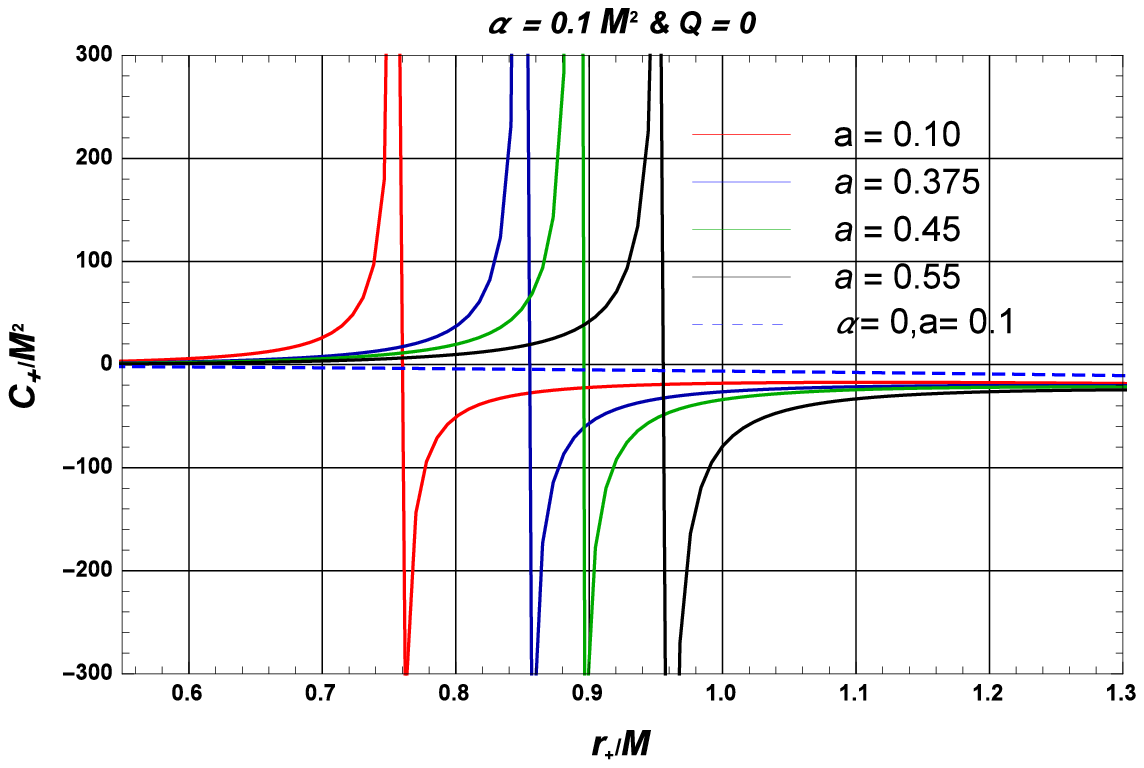}
		\includegraphics[width=0.5\linewidth]{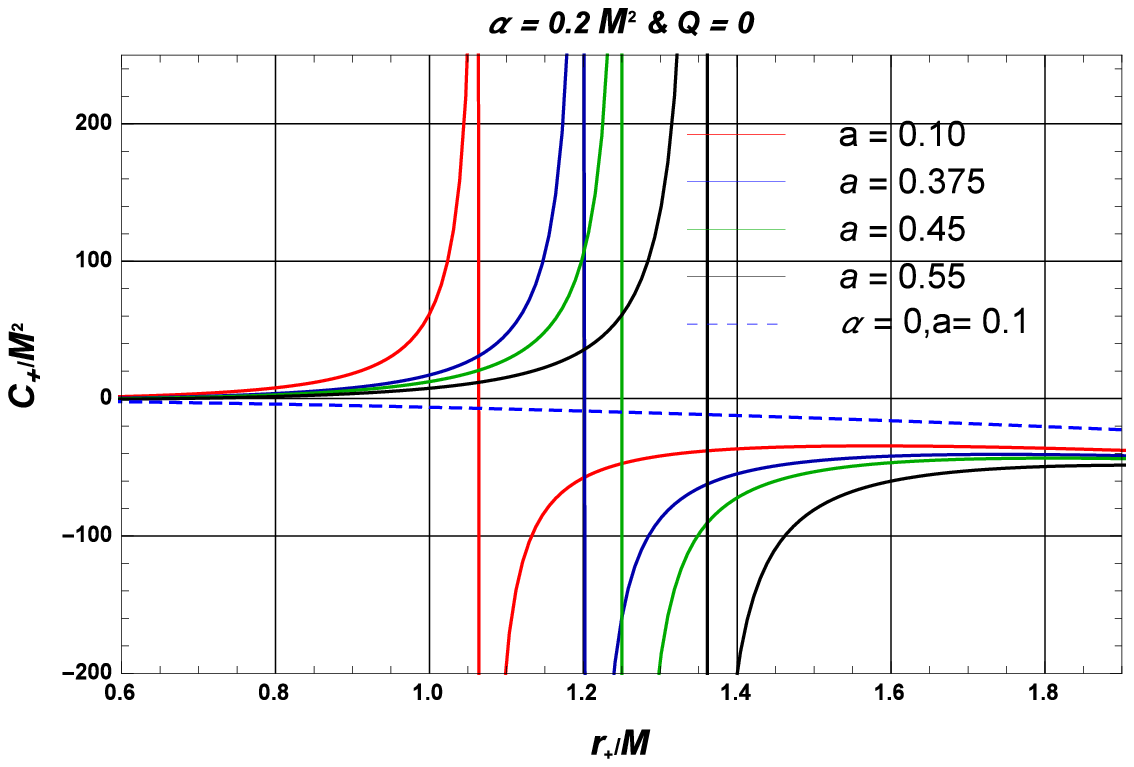}\\
\includegraphics[width=0.5\linewidth]{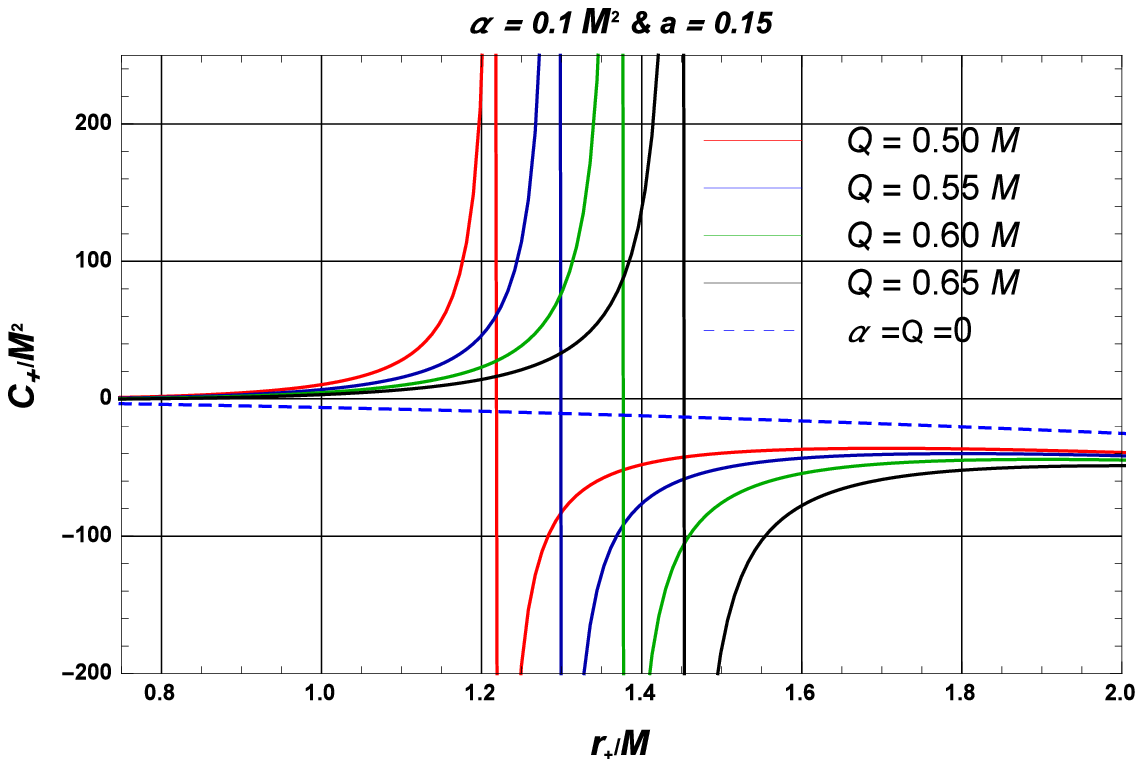}
\includegraphics[width=0.5\linewidth]{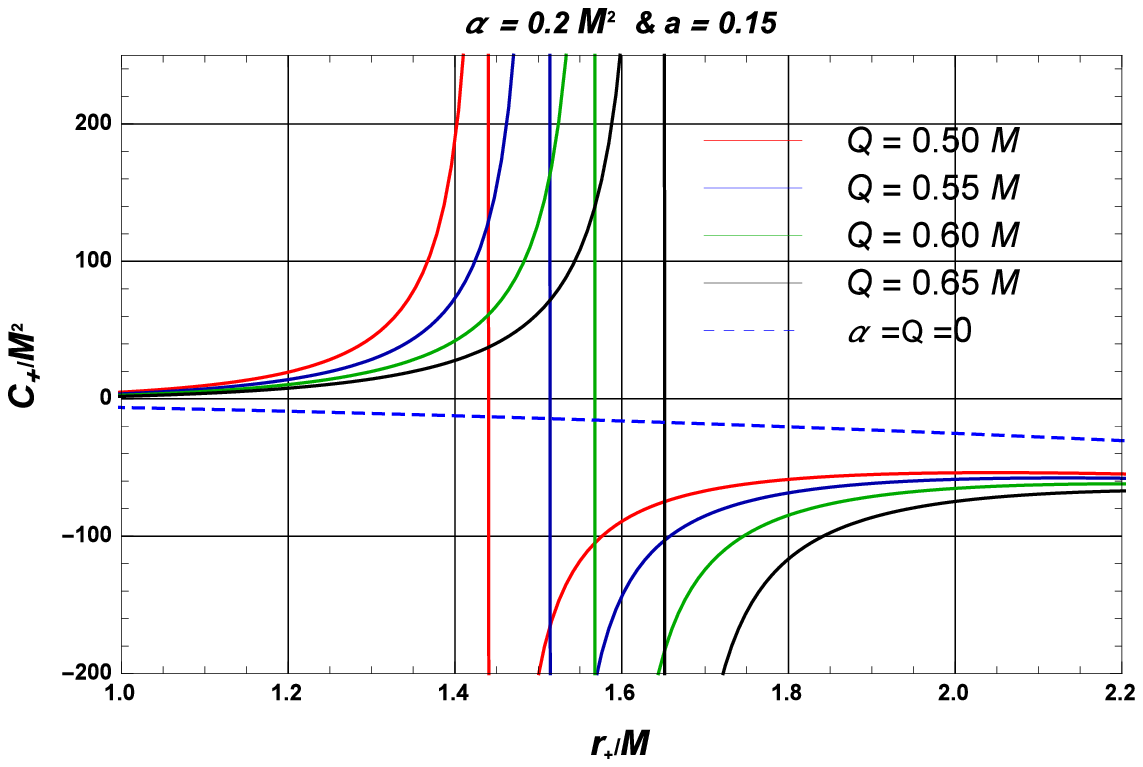}
	\end{tabular}
	\caption{ Plot of heat capacity $C_+$  vs horizon radius $r_+$  both neutral (upper) and charged (lower) $4D$  EGB black hole surrounded by CS.  }
	\label{fig:sh}
\end{figure*}
To further analyse,  we plot the  heat capacity in Fig. \ref{fig:sh} for different values of CS  parameter $a$ and GB coupling constant  $\alpha$,  which clearly exhibits that the heat capacity, for a given value of $a$ and $\alpha$, is discontinuous exactly at the critical radius $r_c$. Further, we note that there is a flip of sign in the heat capacity around $r_c$ . Thus, $4D$  EGB black holes (both charged and uncharged) are  thermodynamically stable for  $r_+ < r_c$, whereas it is thermodynamically unstable for $r_+>r_c$, and there is a phase transition at $r_+=r_c$ from the stable to unstable phases. Further, a divergence of the  heat capacity   at critical $r_+=r_C$ signals a second order phase transition occurs  \cite{hp,davis77}. The heat capacity is discontinuous at $r_+=2.013$, at which the Hawking temperature  has the maximum value $T_+=0.0243$ for $\alpha=0.1$ and $a=0.0909$  (Fig. \ref{fig:sh}). The phase transition occurs from the higher to lower mass black holes corresponding to  negative to positive heat capacity. The critical radius $r_C$ increases with  $M$ (cf. Fig. \ref{fig:sh}), and decreases with $Q$. The stable phase can be seen at the large value of CS parameter $a$ (see also, Table \ref{tab:temp}).

We finally  comment on the black hole remnant  which  is  a source for dark energy \cite{jh}  and also one of the candidates to resolve the information loss puzzle \cite{jp}. The double root $r=r_c$ of $f(r)=0$  corresponds to the extremal black hole with degenerate horizon. Hence
\begin{equation}
f'(r_c)=0 = T_{+},
\label{fr}
\end{equation}
on substituting $f(r)$ from Eq. (\ref{sol1}) to (\ref{fr}), we obtain the critical radius as  (\ref{Rc})
 corresponding to the critical mass  (\ref{Mc}), We can see clearly that  the two horizons  coincide when $r_C\equiv r_-=r_+$ and the temperature decreases with increasing $r_{-}$ and vanishes. Thus we find that the  temperature vanishes at the degenerate horizon leaving a regular double-horizon remnant with $M=M_c$, i.e.,  the black hole
 evaporation are remnants whose near horizon geometry is similar to the extremal black hole geometry with the Hawking temperature of these remnants is equal to  zero. 

\section{Conclusion}
EGB gravity is a natural extension of  Einstein's GR to HD ($D\geq5$) that has several additional nice properties than  GR  and is the first nontrivial term of low energy limit of string theory.  But EGB gravity is topological in $D\leq4$ and does not make a contribution to the gravitational dynamics.  This has been addressed in the $4D$ EGB gravity in which the quadratic GB term in the  action makes a non-trivial contribution to the gravitational dynamics in $4D$ and in contrast to the  Schwarzschild black hole solution of GR, a black hole in this theory is free from the singularity pathology.  However, the quadratic curvature in the theory causes complications in the calculation and hence, in general, investigation of this theory is a bit tedious.  However, later a geodesic analysis contradicts this  observation about the singularity being unreachable by any observer in finite proper time \cite{Arrechea:2020evj}.

Hence,  we have obtained an exact $4D$ static spherically symmetric black hole solution surrounded by the CS to the $4D$ EGB  theory utilizing  the procedure proposed by Glavan and Lin \cite{gla}. However, our spherically symmetric $4D$  black hole solution remains valid in other regularized theories \cite{Lu:2020iav,Hennigar:2020lsl,Casalino:2020kbt}.  It encompasses the known black holes of Glavan and Lin \cite{gla} and Fernandes \cite{Fernandes:2020rpa} of the $4D$ EGB  theory.  In turn,  we have analyzed thermodynamics of the  $4D$  charged EGB black hole with CS to calculate exact expressions for the thermodynamic quantities like the black hole mass, Hawking temperature, entropy, specific heat and analyzed the thermodynamical stability of black holes. The thermodynamical quantities get corrected owing to the background CS, except for the entropy which does not depend on the background CS. The entropy of a black hole has the logarithmic correction to the Bekenstein-Hawking  area law.   The heat capacity increase indefinitely at critical horizon radius $r_+^C$, which depends on both GB coupling constant $\alpha$  and   CS parameter $a$,  where the black hole is extremal and incidentally  local maxima of the Hawking temperature also occur at $r_+^C$, and  that the heat capacity is positive for $r<r_+^C$ implying the stability of small black holes against perturbations in the region, and the phase transition exists at $r_+^C$.  While the black hole is unstable for $r>r_+^C$ with negative heat capacity. Further,  the smaller black hole are globally stable with positive heat capacity $C_+>0$ and negative free energy $F_+<0$. Finally, we have also shown that the black hole evaporation results in a  stable black hole remnant with zero temperature $T_+=0$ and positive specific heat $C_+>0$.   It would be important to understand how these black holes with positive specific heat ($C>0$) would emerge from thermal radiation through a phase transition.    

There are many interesting avenues that are amenable for future work, it will be intriguing to analyse accretion onto the black holes. Since, we find that the background CS makes profound influence as the horizon radius of the black hole under consideration becomes larger which may have several astrophysical consequences, like on wormholes, gravitational lensing and black holes in AdS/CFT.  Some of the results presented here are generalization of the previous discussions, on the $4D$ EGB  \cite{gla, Fernandes:2020rpa} and GR black holes \cite{Ghosh:2014pga}, in a more general setting, and the possibility of a further generalization of these results to Lovelock gravity \cite{Konoplya:2020qqh} is an interesting problem for future.  One can also think, in the spirit of the no-hair conjectures \cite{wheeler}, how two different matters viz. CS and global monopole can generate the same spacetime (\ref{sol1}) or have same stress-energy tensor.

\acknowledgments
Authors would like to thank DST INDO-SA bilateral project DST/INT/South Africa
/P-06/2016, S.G.G. also thank SERB-DST for the ASEAN project IMRC
/AISTDF/CRD/2018/ 000042, and Rahul Kumar for help in plots and fruitful discussion. S.D.M. acknowledges
that this work is based upon research supported by the South African
Research Chair Initiative of the Department of Science and
Technology and the National Research Foundation.

\end{document}